%% file: main.tex
\documentclass[letterpaper, 10pt, conference]{ieeeconf}

\IEEEoverridecommandlockouts
\usepackage{amsmath,amssymb,mathtools}
\usepackage{algorithm}
\usepackage{algpseudocode}
\usepackage{booktabs} 
\usepackage{cases}
\usepackage{color}
\usepackage[hidelinks]{hyperref}
\usepackage{lipsum} 
\usepackage{microtype}

\usepackage[short]{optidef} 
\usepackage{siunitx} 
\usepackage{statmath}
\usepackage{subcaption}
\usepackage{svg}
\usepackage{tabularx} 
\usepackage{tikz}
\usetikzlibrary{arrows, arrows.meta, calc, fit, positioning, shapes.geometric}
\usepackage{wrapfig}

\newtheorem{proposition}{Proposition}
\newtheorem{remark}{Remark}

\newcommand{\R}{\mathbb{R}}
\newcommand{\Aset}{\mathcal{A}}
\newcommand{\Iset}{\mathcal{I}}
\newcommand{\Fset}{\mathcal{F}}
\newcommand{\Flin}{\mathcal{F}_{\mathrm{lin}}}

\newcommand{\Xfeas}{\mathcal{X}_{0}}
\newcommand{\Xlin}{\mathcal{X}_{\mathrm{lin}}}

\begin{document}

\title{Computationally Efficient Data Generation for Learning \\ Nonlinear MPC Policies}

\author{Pietro Gori$^{1}$, Anton Pozharskiy$^{2}$, Franco Angelini$^{1}$, Manolo Garabini$^{1}$, Moritz Diehl$^{2}$, and Armin Nurkanovi\'c$^{2}$
\thanks{This research is partially supported by the Italian Ministry of University and Research (MUR) - Fondo Italiano per la Scienza Applicata (FISA), through the OCCAM project under Grant FISA 2023-00324 CUP I53C25000700001, partially by the European Union HORIZON-MSCA-2023-SE-01-01 - MSCA Staff Exchanges 2023 Program under the Grant Agreement No.101182891 (NEUTRAWEED), partially by ``FoReLab'' (Future-oriented Research Lab) Project (Departments of Excellence), and partially by DFG via projects 504452366 (SPP 2364), 560056112 (robust MPC), 535860958 (ALeSCo) and 525018088 (MAWERO), and by BMWK via 03EN3054B}
\thanks{$^{1}$Centro di Ricerca `Enrico Piaggio'' and the Dipartimento di Ingegneria dell'Informazione,  Universita' di Pisa, Italy (email: {\tt\footnotesize pietro.gori@phd.unipi.it}).}
\thanks{$^{2}$Department of Microsystems Engineering (IMTEK), University of Freiburg, Germany}
}

\maketitle

\begin{abstract} 
\input{sections/abstract}
\end{abstract}


\IEEEpeerreviewmaketitle


\input{sections/introduction}

\input{sections/problem_definition}

\input{sections/active_set_exploration_strategies}

\input{sections/projection_based_generation}

\input{sections/examples}

\input{sections/conclusions}



\bibliographystyle{IEEEtran}
\bibliography{Bibliography}

\end{document}

%% file: sections/abstract.tex
Learning MPC requires large datasets generated by repeatedly solving MPC problems. 
For Nonlinear MPC (NMPC), dataset generation is computationally expensive because feasible initial states are difficult to sample efficiently.
The feasible set is often non-convex or even disconnected, and many randomly sampled states are infeasible, leading to wasted optimization solves and unnecessary computational effort.
In this work, we propose a data-generation framework for learning NMPC policies that combines active-set exploration with projection-based sampling. 
First, a bilevel formulation samples activity weights to promote solutions associated with different active sets while preserving optimality for the original NMPC problem; second, a simpler slack-based formulation uses weighted slack variables to target different active sets at the single level. 
These reference solutions are then used to construct local linearized feasible sets, from which candidate initial states are sampled and projected through the nonlinear NMPC problem to recover feasible and locally optimal training samples.
Our approach is up to $37\times$ faster than baseline uniform sampling with $99\%$ acceptance rate, while maintaining a representative coverage of the feasible set.

%% file: sections/introduction.tex
\section{Introduction}
\label{sec:introduction}
Machine-learning-based control has emerged as a promising approach for deploying advanced control strategies on systems with limited computational resources. 
In particular, learning NMPC aims to replace the online solution of an NMPC problem with evaluating a neural network (NN), 
which is computationally cheaper, that approximates the NMPC control law \cite{lucia2020deep,wang2023policy}. 
%
The performance of the approximating NN policy depends critically on the quality of the training dataset. 
A common approach is to generate this dataset by repeatedly solving the NMPC problem for a large number of sampled initial states. 
For constrained nonlinear systems, however, this process can be highly inefficient because many randomly sampled states are infeasible, 
leading to wasted optimization solves and unnecessary computational effort. 
This challenge becomes particularly pronounced in problems involving complex nonlinear dynamics and nonlinear, non-convex constraints, 
for which the generation of labeled data often constitutes the primary computational bottleneck.
Consequently, efficient sampling strategies are essential for constructing high-quality training datasets. 
Rather than increasing the number of samples, the objective is to generate informative and feasible samples 
that adequately represent the domain of interest while minimizing the computational effort required for dataset generation.

In this work, we propose two active set exploration strategies and a projection-based data-generation method for efficiently generating feasible initial states for learning NMPC policies.
Those strategies can be combined to further enhance data-generation efficiency. 
The proposed methods substantially reduce the computational cost of dataset generation by avoiding unnecessary optimization solves 
while preserving a representative coverage of the feasible region.

\input{sections/related_works}

%% file: sections/related_works.tex
\section{Related Works}
\label{sec:related_works}
Common approaches for learning NMPC policies train a neural network or other function approximator on state--input pairs, where the input is the first element of the optimal control sequence returned by the NMPC problem. 
Although this shifts most computation offline, dataset generation remains expensive because representative states must cover a feasible set that is typically known only implicitly through problem solvability. 
Since each labeled sample generally requires solving a constrained Optimal Control Problem (OCP), the sampling strategy directly affects both offline cost and controller quality. 
Efficient initial-state sampling is also relevant in explicit MPC, where feasible-set exploration helps identify distinct active-set configurations~\cite{kenefake2024novel}.

Several works have addressed feasible-state sampling for Linear MPC (LMPC). 
Winqvist et al.~\cite{winqvist2020training} proposed a Hit-and-Run method for efficiently sampling polyhedral feasible sets, while Chen et al.~\cite{chen2022large} generated feasible datasets through directional random walks and repeated quadratic-program solves. 
More recently, Milios et al.~\cite{milios2026efficient} proposed LMPC-HR, which identifies feasible chord boundaries through linear programs and enables approximately uniform sampling at lower computational cost. 
These approaches exploit the convex polyhedral structure of LMPC feasible sets and do not readily extend to NMPC.

Sampling is more challenging for NMPC, where the feasible set is available only implicitly and may be non-convex or disconnected. 
Johansen~\cite{johansen2004approximate} used adaptive gridding over a prescribed state domain, while Chakrabarty et al.~\cite{chakrabarty2016support} employed deterministic space-filling samples to learn approximations of the feasible region and control law. 
More recently, Tokmak et al.~\cite{tokmak2025automatic} proposed an adaptive kernel interpolation method that refines a prescribed domain according to approximation error. 
These methods improve sample allocation over a chosen domain but do not directly address efficient exploration of an implicitly defined NMPC feasible set.

Complementary approaches improve the information obtained from each expensive NMPC solution. 
Sensitivity-based augmentation~\cite{krishnamoorthy2021sensitivity,krishnamoorthy2023improved}, gradient-enhanced learning~\cite{winqvist2021learning}, and Lipschitz-guided dataset design~\cite{alsmeier2024imitation} exploit local model information to reduce the number of required exact solves. 
Other methods instead prioritize closed-loop relevance, for example by sampling around nominal trajectories or according to approximation error~\cite{tagliabue2024efficient,zamora2022learning}. 
These strategies improve local or task-specific sample efficiency but do not aim at systematic exploration of the entire NMPC feasible set.

In summary, existing methods either exploit convex LMPC structure, sample or refine a prescribed state domain, or improve label efficiency around previously selected states. 
To the best of our knowledge, no existing approach specifically targets efficient generation of representative feasible initial states for NMPC problems with implicitly defined, possibly non-convex feasible sets.
\subsection{Contributions}


In this work, we propose an efficient procedure for generating feasible initial states for learning NMPC policies by combining active-set exploration with projection-based sampling. 
The main contributions of this work are:
\begin{enumerate}
    \item We introduce two active-set exploration strategies for generating feasible initial states: a bilevel formulation that promotes different active sets while preserving optimality for the original NMPC problem, and a simpler formulation based on weighted slack variables. We characterize the optimality properties of both.

    \item We develop a projection-based sampling strategy that draws candidates from a linearized feasible set and projects them through a nonlinear OCP. We characterize when the linearization is an inner or outer approximation and show that the projection returns feasible, locally optimal solutions for the corresponding original NMPC problem.

    \item We evaluate the combined strategies on nonlinear MPC examples, assessing acceptance rate, computational cost, and coverage of the feasible set against uniform sampling and closed-loop rollouts.
\end{enumerate}
The proposed approach is up to $37 \times$ faster than baseline uniform sampling, achieves acceptance rates above $99\%$, and maintains representative coverage of the feasible set.

%% file: sections/problem_definition.tex
\section{Problem Definition}
\label{sec:problem_definition}

\subsection{Notation}
Throughout the paper, the concatenation of vectors is denoted by $(a,b):=[a^\top, b^\top]^\top$, and analogously for more than two vectors.
Furthermore, we use $\nabla f(w)$ for the gradient of the scalar-valued function and $\nabla_w h(w):=\frac{\partial h(w)}{\partial w}^\top$,
for the Jacobians of the vector-valued constraint functions.
For a positive semidefinite matrix $Q\succeq 0$, we use the weighted norm notation $\|s\|_Q^2:=s^\top Qs$.
Finally, for an index set $\mathcal{I}$, $x_{\mathcal{I}}$ denotes the subvector of $x$ corresponding the components indexed by $\mathcal{I}$, i.e., $x_{\mathcal{I}} := \{x_i \mid i \in \mathcal{I}\}$.
\subsection{Discrete-Time Optimal Control Problem}
For a prescribed initial state $\bar x_0$, consider a discrete-time nonlinear OCP
\begin{mini!}|s|[2]
    {\substack{x_0,\ldots,x_N\\u_0,\ldots,u_{N-1}}}
    {\displaystyle \sum_{k=0}^{N-1}L_k(x_k,u_k)+V_{\mathrm{f}}(x_N)}
    {\label{eq:ocp_explicit}}{}
    \addConstraint{x_0}{=\bar x_0}
    \addConstraint{x_{k+1}}{=F_k(x_k,u_k), \quad}{k=0,\ldots,N-1}
    \addConstraint{g_k(x_k, u_k)}{\leq0,}{k=0,\ldots,N-1,}
\end{mini!}
where $x_k \in \R^{n_x}$ and $u_k \in \R^{n_u}$ are the state and control input at a generic time step $k$, respectively.
The function $L_k: \R^{n_x} \times\R^{n_u} \to \R$ is the stage cost, $V_{\mathrm{f}}:\R^{n_x} \to \R$ is the terminal cost, $F_k:\R^{n_x} \times \R^{n_u} \to \R^{n_x}$ is the state transition map, and $g_k:\R^{n_x} \times \R^{n_u} \to \R^{m}$ is the vector of inequality constraints.
The horizon length is denoted by $N$.

Our goal is to find initial states $\bar x_0$ feasible for \eqref{eq:ocp_explicit} that lead to optimal trajectories with different active sets.
To this end, we first rewrite \eqref{eq:ocp_explicit} in a more compact form.
We collect all degrees of freedom in $w:=~(x_0,x_1,\ldots,x_N,u_0,\ldots,u_{N-1})=(x_0,y)$,
where $y:=(x_1,\ldots,x_N,u_0,\ldots,u_{N-1})$.
Hence, we define $f(w):=\sum_{k=0}^{N-1}L_k(x_k,u_k)+V_f(x_N)$,	and the vector-valued functions $g,h$ collect the inequality and equality constraints, respectively. 
Thus, we rewrite \eqref{eq:ocp_explicit} compactly as:
\begin{mini!}|s|[2]
    {\substack{w}}{f(w)}
    {\label{eq:ocp_compact}}{}
    \addConstraint{h(w,\bar x_0)}{=0}
    \addConstraint{g(w)}{\leq0.}
\end{mini!}
We note that the parameter $\bar x_0$ appears only in the equality constraint $h(w,\bar x_0)=0$. 
Denote the optimization problem in \eqref{eq:ocp_compact} by $\mathcal P(\bar x_0)$, and define the set of feasible initial states as
\[
    \Xfeas:=
    \left\{
        \xi\in\R^{n_x}\ \middle|\
        \exists w:\ h(w,\xi)=0,\ g(w)\leq0
    \right\}.
\]

Our objective is to identify elements of \(\Xfeas\) whose corresponding solutions to \(\mathcal P(\bar x_0)\) realize distinct active sets. 
To this end, we formulate auxiliary optimization problems that treat the initial state as an implicit decision variable.

%% file: sections/active_set_exploration_strategies.tex
\section{Active Set Exploration Strategies}
\label{sec:active_set_exploration}
This section presents two formulations for generating feasible initial states while exploring different NMPC active sets. 
Both approaches rely on sampled weights to steer the search toward different constraint-activation patterns. 
We first introduce a bilevel formulation that preserves optimality with respect to the original NMPC problem, 
followed by a single-level formulation based on slack variables.

\subsection{Bilevel Active-Set Selection}
\label{sec:bilevel_active_set_selection}
We present a bilevel optimization formulation for generating initial states associated with different active sets. 
By construction, the generated initial states are feasible, and their corresponding solutions satisfy the Karush-Kuhn-Tucker (KKT) conditions of \eqref{eq:ocp_explicit}. 
Under appropriate convexity assumptions, these solutions are also globally optimal.

We initially introduce the slack variables $s\in\R_{\geq0}^m$ necessary to reformulate the inequality constraints in \eqref{eq:ocp_compact} as $g(w)+s=0$,
so that $s_i=0$ if and only if the $i$-th inequality is active, and $s_i >0$ implies $g_i(w) < 0$, i.e., the constraint is inactive.
In this way, the active set of a feasible $w$ can be determined by the corresponding slack variable $s$.

Let $Q=\operatorname{diag}(q_1,\ldots,q_m)\succeq 0$ collect the sampled activity weights, and define $\mathcal T:=\{i\mid q_i>0\}$ as the corresponding set of targeted inequalities. 
To avoid overconstraining the problem, the number of simultaneously targeted constraints should, as a dimension-counting guideline, satisfy $m_h+|\mathcal T|\leq n_w$,
where $m_h=\dim h(w)$ and $n_w=\dim w$. 
This prevents the resulting active-set system from being generically overdetermined. 
For lower and upper bounds on the same scalar variable, at most one bound should be assigned a positive weight in each sample, unless the bounds coincide. 
An optional sampling cost $r(x_0)$ can further guide the selection of the generated initial state, for instance, $r(x_0)= \beta\|x_0-\hat{x}_0\|_2^2$, where $\hat{x}_0$ is a reference state and $\beta>0$ is a weighting factor.
Feasible $x_0 \in \Xfeas$, with different active sets, can be found by solving the following Mathematical Program with Equilibrium Constraints (MPEC):
\begin{mini!}|s|[2]
 {\substack{x_0,w,s}}
 {r(x_0)+\frac{1}{2}\|s\|_Q^2}
 {\label{eq:bilevel}}{}
 \addConstraint{g(w)+s}{=0}
 \addConstraint{s}{\geq0}
 \addConstraint{w}{\in\operatorname*{argmin}_{\widetilde w}
 \left\{f(\widetilde w)\ \middle|\
 h(\widetilde w, x_0)=0,\ g(\widetilde w)\leq0\right\}.}
\end{mini!}
\vspace{-0.3cm}
\begin{remark}
We formulate the results using a weighted Euclidean norm for the slack variables.
The same analysis carries over to a weighted $\ell_1$ norm by replacing $\frac{1}{2}\|s\|_Q^2$ with $q^\top s$.
The statements and proofs of the propositions below require only minor adjustments.
\end{remark}


Following the KKT reformulation strategy of \cite{luo1996mathematical}, the lower-level MPC problem is replaced by its parametric KKT conditions, 
yielding a single-level Mathematical Program with Complementarity Constraints (MPCC). 
Let $\lambda$ and $\mu$ be the lower-level multipliers associated with $h$ and $g$, respectively. 
Replacing the lower-level problem by its KKT conditions gives the following MPCC:
\begin{mini!}|s|[2]
 {\substack{x_0, w,s,\lambda,\mu}}
 {r(x_0)+\frac12\|s\|_Q^2}
 {\label{eq:mpcc}}{}
 \addConstraint{g(w)+s}{=0}
 \addConstraint{h(w,x_0)}{=0}
 \addConstraint{\nabla_w f(w)+\nabla_w h(w,x_0)\lambda+\nabla_w g(w)\mu}{=0}
 \addConstraint{0\leq\mu}{\perp s\geq0.}
\end{mini!}
%
MPCCs are generally more challenging to solve than standard NLPs because their complementarity constraints violate standard constraint qualifications at feasible points. 
Nevertheless, dedicated reformulations and numerical methods have enabled efficient solution of large-scale MPCCs~\cite{kim2020mpec,nurkanovic2025solving,pozharskiy2026ccopt}.

For a nonconvex lower-level problem, the KKT conditions are generally only necessary for local optimality under an appropriate constraint qualification. 
Consequently, \eqref{eq:mpcc} constitutes a relaxation of the bilevel formulation and may admit lower-level KKT points that are not local minimizers. 
In contrast, when the lower-level problem is convex and satisfies suitable regularity conditions, the KKT conditions are also sufficient for global optimality, and the reformulation is exact~\cite{luo1996mathematical}.

By solving \eqref{eq:mpcc} for different sampled weights $Q$, we find feasible initial states $x_0\in\Xfeas$ with different active sets, whose corresponding solutions $w$ satisfy the KKT conditions of \eqref{eq:ocp_compact}.


\subsection{Single-Level Slacked Formulation}
\label{sec:single_level_slacked}
As an alternative to the bilevel formulation, we find feasible initial states associated with different active sets by introducing slack variables into the original NMPC problem.
Consider the following slacked formulation
\begin{mini!}|s|[2]
	{\substack{w,s}}
	{f(w)+\frac12\|s\|_Q^2}
	{\label{eq:slack_problem}}{}
	\addConstraint{\tilde h(w)}{=0}
	\addConstraint{g(w)+s}{=0}
	\addConstraint{s}{\geq0,}
\end{mini!}
where $Q=\diag(q_1,\ldots,q_m)\succeq0$ contains the sampled activity weights.
Since $g(w)=-s\leq0$, every feasible solution $w=~(x_0,y)$ of \eqref{eq:slack_problem} is feasible for the original MPC problem $\mathcal P(x_0)$.
The expression $\tilde h(w)$ is obtained from $h(w,x_0)$ by dropping the constraints that fix $x_0$, so that $x_0$ becomes a degree of freedom.
Similar to \eqref{eq:mpcc}, positive entries of $Q$ penalize the corresponding slacks, steering the solution toward states at which the associated constraints are active.
The number of simultaneously targeted constraints should satisfy the same dimension-counting guideline as in the bilevel formulation, i.e., $m_h+|\mathcal T|\leq n_w$, to avoid overconstraining the problem.

We next characterize the effect of the slack-based sampling objective on the optimality conditions of the corresponding original MPC problem.
Let $\lambda$, $\pi$, and $\nu\geq0$ denote the multipliers associated with $h(w)=0$, $g(w)+s=0$, and $-s\leq0$, respectively.
The KKT conditions of \eqref{eq:slack_problem} are
\begin{subequations}
\begin{align}
    \nabla f(w) + \nabla h(w)\lambda + \nabla g(w)\pi &= 0, \label{eq:kkt_w} \\
    Qs + \pi - \nu &= 0, \label{eq:kkt_s} \\
    h(w) = 0, \qquad g(w) + s &= 0, \label{eq:kkt_feas} \\
    0 \leq \nu \perp s &\geq 0. \label{eq:kkt_comp}
\end{align}
\end{subequations}

At a stationary point $(w^*,s^*)$, define the active and inactive index sets as $\Aset:=\{i\mid s_i^*=0\}$, $\Iset:=\{i\mid s_i^*>0\}$.
Since $Q$ is diagonal, \eqref{eq:kkt_s}--\eqref{eq:kkt_comp} yield $\pi_{\Aset}^*=\nu_{\Aset}^*\geq0$, $\pi_{\Iset}^*=-Q_{\Iset}s_{\Iset}^*$.
Consequently, the $y$-component of the stationarity condition \eqref{eq:kkt_w} becomes
\begin{equation}
\begin{aligned}
\nabla_y f(w^*)
+\nabla_y h(w^*)\lambda^*
+\nabla_y g_{\Aset}(w^*)\pi_{\Aset}^*
&\\
-\nabla_y g_{\Iset}(w^*)
Q_{\Iset}s_{\Iset}^*
&=0.
\end{aligned}
\label{eq:y_stationarity}
\end{equation}
Define the perturbation induced by the sampling objective as
\begin{equation}
c^*:= -\nabla_y g_{\Iset}(w^*)
Q_{\Iset}s_{\Iset}^*.
\label{eq:perturbation}
\end{equation}
Then \eqref{eq:y_stationarity} can be written as 
\begin{equation}
\nabla_y f(w^*)+c^*
+\nabla_y h(w^*)\lambda^*
+\nabla_y g_{\Aset}(w^*)\pi_{\Aset}^*
=0.
\label{eq:perturbed_stationarity}
\end{equation}
Hence, the additional stationarity contribution introduced by the slack penalty admits a direct interpretation as a linear perturbation of the original NMPC objective.

\begin{proposition}[Equivalent perturbed MPC problem]
\label{prop:perturbed_mpc}
Let $(w^*,s^*)$, with $w^*=(x_0^*,y^*)$, be a KKT point of
\eqref{eq:slack_problem}, with corresponding multipliers
$(\lambda^*,\pi^*,\nu^*)$, and define $c^*$ according to
\eqref{eq:perturbation}.
Then $w^*$ is a KKT point of the perturbed MPC problem
\begin{mini!}|s|[2]
	{\substack{w}}
	{f(w)+(c^*)^\top y}
	{\label{eq:perturbed_mpc}}{}
	\addConstraint{h(w,x_0^*)}{=0}
	\addConstraint{g(w)}{\leq0.}
\end{mini!}
In particular, if $c^*=0$, then $w^*$ is a KKT point of the original MPC problem $\mathcal P(x_0^*)$.
\end{proposition}

\begin{proof}
Primal feasibility of \eqref{eq:perturbed_mpc} follows immediately from
\eqref{eq:kkt_feas}, since $g(w^*)=-s^*\leq0$.
Define the inequality multipliers of the perturbed MPC problem as $\mu_{\Aset}^*:=\pi_{\Aset}^*\geq0$, $\mu_{\Iset}^*:=0$.
Since $g_{\Aset}(w^*)=0$, $g_{\Iset}(w^*)<0$, these multipliers satisfy dual feasibility and complementarity, $0\leq\mu^*\perp-g(w^*)\geq0$.

By the definition of $c^*$ in \eqref{eq:perturbation},
the $y$-component stationarity condition
\eqref{eq:y_stationarity} becomes \eqref{eq:perturbed_stationarity}.

It remains to recover stationarity with respect to $x_0$.
Let $\alpha^*$ denote the multiplier associated with the initial-value constraint $x_0-x_0^*=0$ in \eqref{eq:perturbed_mpc}, and choose
\[
\alpha^*
:=
-\nabla_{x_0}f(w^*)
-\nabla_{x_0}h(w^*)^\top\lambda^*
-\nabla_{x_0}g_{\Aset}(w^*)^\top\mu_{\Aset}^*.
\]
Since the perturbation $(c^*)^\top y$ does not depend on $x_0$, no additional term appears in this expression.
Then, with $\hat\lambda^*:=(\alpha^*,\lambda^*)$, the complete stationarity condition of the perturbed MPC problem is
\[
\nabla f(w^*)
+\bar c^*
+\nabla_w h(w,x_0^*)\big|_{w=w^*}^{\top}\hat\lambda^*
+\nabla_w g(w^*)^\top\mu^*
=0.
\]
Together with primal feasibility, dual feasibility, and complementarity, this establishes the KKT conditions of \eqref{eq:perturbed_mpc}.

If $c^*=0$, the perturbed objective coincides with the original MPC objective, and therefore $w^*$ is a KKT point of $\mathcal P(x_0^*)$.
\end{proof}

\begin{remark}
The vector $c^*$ quantifies the objective perturbation induced by the slack-based sampling criterion.
In particular, $\|c^*\| = \left\| \nabla_y g_{\Iset}(w^*)^\top Q_{\Iset}s_{\Iset}^* \right\|$
provides an a posteriori measure of the deviation from stationarity of the original MPC problem.
A vanishing perturbation, $c^*=0$, implies exact KKT compatibility with $\mathcal P(x_0^*)$.
Importantly, this condition does not require each individual weighted slack $q_i s_i^*$ to vanish: contributions associated with different inactive constraint gradients may cancel in \eqref{eq:perturbation}.
\end{remark}

%% file: sections/projection_based_generation.tex
\section{Projection-based Sampling}
\label{sec:projection_based_sampling}
The active set exploration strategies allow us to promote different active sets.
However, their sampled states may perturb the optimality conditions of the original MPC problem.
We therefore consider a complementary projection-based strategy in which a reference solution $\hat w$ is first obtained using one of the active-set exploration strategies in Sec.~\ref{sec:active_set_exploration}.
To enable efficient sampling, we approximate the nonlinear feasible set around $\hat w$ by a tractable linearized set $\Xlin(\hat w)$, from which candidate initial states are sampled~\cite{milios2026efficient}.
Each candidate $\bar x_0 \in \Xlin(\hat w)$ is then projected onto the nonlinear feasible set $\Xfeas$ by solving the following OCP:
\begin{mini!}|s|[2]
    {\substack{w}}
    {f(w)+\frac12 \|x_0-\bar x_0\|_{Q_x}^2}
	{\label{eq:projection_problem}}{}
    \addConstraint{\tilde h(w)}{=0}
    \addConstraint{g(w)}{\leq0,}
\end{mini!}
where $Q_x\succ0$.
Recall that $\tilde h(w)$ is obtained from $h(w,\bar x_0)$ by removing the constraint that fixes $x_0$.
The additional term in the objective penalizes the distance between the generated initial state $x_0$ and the target $\bar x_0$.
Since it only depends on $x_0$, if $w^*=(x_0^*,y^*)$ is a local minimizer of \eqref{eq:projection_problem},
then it is also a local minimizer of $\mathcal P(x_0^*)$, because the projection term is constant when $x_0=x_0^*$ is fixed.

Let $\hat w=(\hat x_0,\hat y)$ be a feasible reference solution, i.e., $\tilde h(\hat w)=~0$, $g(\hat w)\leq0$.
We consider the free-initial-state feasible set
\[
\Fset := \left\{ w=(x_0,y) \;\middle|\; \tilde h(w)=0,\; g(w)\leq0 \right\}.
\]

A first-order approximation of $\Fset$ around $\hat w$ is obtained by
linearizing the equality and inequality constraints. Writing
$w=\hat w+ \Delta w$, we define
\begin{equation*}
\begin{aligned}
\mathcal{F}_{\mathrm{lin}}(\hat w) := \bigl\{\, \hat w+\Delta w \;\big|\;& \nabla_w \tilde h(\hat w)^\top \Delta w=0,\\
&g(\hat w)+\nabla_w g(\hat w)^\top \Delta w\leq0 \,\bigr\}.
\end{aligned}
\end{equation*}
Here, we used $\tilde h(\hat w)=0$ since $\hat w$ is feasible.

Since we are interested in sampling initial states rather than complete
trajectories, we consider the projection of $\Flin(\hat w)$ onto the
initial-state coordinates. Partitioning $\Delta w=(\Delta x_0,\Delta y)$, the corresponding linearized feasible set of initial states is
\begin{equation*}
\begin{aligned}
\mathcal{X}_{\mathrm{lin}}(\hat w) := \bigl\{\, \hat x_0+\Delta x_0 \;\big|\;& \exists\,\Delta_y \text{ such that}\\
&\hat w+(\Delta x_0,\Delta y) \in\mathcal{F}_{\mathrm{lin}}(\hat w) \,\bigr\}.
\end{aligned}
\end{equation*}
The set $\Xlin(\hat w)$ therefore contains all initial states that admit a
feasible trajectory in the linearization of the OCP constraint functions around $\hat w$.
To obtain a bounded sampling domain, the linearized feasible set is
intersected with a prescribed local sampling region.
Candidate initial states $\bar x_0$ are then sampled from this bounded
linearized approximation, e.g., using a polyhedral hit-and-run method
\cite{winqvist2020training,milios2026efficient}, and subsequently used as
targets in the nonlinear projection problem \eqref{eq:projection_problem}.

\subsection{Discussion of linearization properties}
A natural question that may arise is whether an initial state $\bar x_0 \in \Xlin(\hat w)$ also belongs to $\Xfeas$.
Before introducing the sampling procedure, it is useful to clarify the relationship between the linearized and nonlinear feasible sets, 
as feasibility for the linearized OCP does not generally imply feasibility for the original nonlinear OCP. 
The following propositions formalize this relationship.
\begin{proposition}[Outer approximation]
\label{prop:outer}
Suppose that $h$ is affine and every component $g_i$ is convex and differentiable. 
Then $\Fset\subseteq~\Flin(\hat w)$ for every feasible reference point $\hat w$. 
Consequently, $\Xfeas~\subseteq\Xlin(\hat w)$.
Thus, for inequalities written as $g(w)\leq0$, the linearized feasible set is an outer approximation.
\end{proposition}

\begin{proof}
Since $h$ is affine, its linearization is exact. 
Convexity of $g_i$ gives $g_i(w)\geq g_i(\hat w)+\nabla g_i(\hat w)^\top(w-\hat w)$.
If $w\in\Fset$, then $g_i(w)\leq0$, and hence $g_i(\hat w)+\nabla g_i(\hat w)^\top(w-\hat w) \leq g_i(w)\leq~0$.
Therefore $w\in\Flin(\hat w)$. 
Projecting both sets onto the initial-state coordinates preserves the inclusion, yielding $\Xfeas \subseteq \Xlin(\hat w)$.
\end{proof}

\begin{proposition}[Inner approximation]
\label{prop:inner}
Suppose that $h$ is affine and every component $g_i$ is concave and differentiable. 
Then $\Flin(\hat w)\subseteq\Fset$, $\Xlin(\hat w)\subseteq\Xfeas$.
\end{proposition}

\begin{proof}
Concavity gives $g_i(w)\leq g_i(\hat w)+\nabla g_i(\hat w)^\top(w-\hat w)$.
Thus every point satisfying the linearized inequality also satisfies $g_i(w)\leq0$. 
The equality constraints are exact because $h$ is affine. 
Projection onto the initial-state component gives the second inclusion.
\end{proof}

\subsection{Boundary identification for linearized feasible sets}
Here, we adapt the boundary-identification method of \cite{milios2026efficient}, originally proposed for linear MPC problems, to efficiently determine the boundaries of the linearized feasible set $\Xlin(\hat w)$.
In contrast to the original Hit-and-Run method \cite{winqvist2020training}, which identifies the feasible interval along a given direction through repeated feasibility queries, the adapted method computes the boundary directly by solving an optimization problem.

To exploit the OCP structure of $\Flin(\hat w)$, let $x:=~(x_1,\ldots,x_N)$, $u:=~(u_0,\ldots,u_{N-1})$,  $w:=~(x_0,x,u)$,
and let $\hat w=~(\hat x_0,\hat x,\hat u)$ be a feasible reference solution.
Define the perturbations $\Delta x_k:=x_k-\hat x_k$, $\Delta u_k:=u_k-\hat u_k$, and the linearized dynamics matrices $A_k:=\nabla_x F(\hat x_k,\hat u_k)^\top$, $B_k:=\nabla_u F(\hat x_k,\hat u_k)^\top$, for $k=0,\ldots,N-1$.
Here, we distinguish state and input constraints, since the former depend on the state perturbations propagated through the linearized dynamics, whereas the latter depend directly on the input perturbations
Define $C_k :=~\nabla_{x} g_x(\hat x_k)^\top$ and $D_k :=~\nabla_{u} g_u(\hat u_k)^\top$, for $k=0,\ldots,N-1$.
Using this notation, the previously defined set $\Flin(\hat w)$ can be expressed through the linearized dynamics 
\[
\Delta x_{k+1}
=
A_k\Delta x_k+B_k\Delta u_k,
\qquad k=0,\ldots,N-1,
\]
and the linearized inequality constraints
\[
\begin{bmatrix}
	g_x(\hat x_k)\\
	g_u(\hat u_k)
\end{bmatrix}
+
\begin{bmatrix}
	C_k & 0 \\
	0 & D_k
\end{bmatrix}
\begin{bmatrix}
	\Delta x_k\\
	\Delta u_k
\end{bmatrix}
\leq
\begin{bmatrix}
	0\\
	0
\end{bmatrix},
\quad k=0,\ldots,N-1.
\]
%
We recursively eliminate the state perturbations using the linearized dynamics, yielding a condensed polyhedral representation of \(\Flin(\hat w)\) in terms of \(\Delta x_0\) and \(\Delta u\). 
This representation is then used in the boundary-identification procedure adapted from \cite{milios2026efficient}. 
The condensed linearized dynamics are
\begin{equation}
	\Delta x=\Omega\Delta x_0+\Gamma\Delta u,
	\label{eq:linearized_condensed_dynamics}
\end{equation}
where
\[
\Omega
:=
\begin{bmatrix}
	A_0\\
	A_1A_0\\
	A_2A_1A_0\\
	\vdots\\
	A_{N-1}A_{N-2}\cdots A_0
\end{bmatrix},
\]
and
\[
\Gamma
:=
\begin{bmatrix}
	B_0 &0 &\cdots &0\\
	A_1B_0 &B_1 &\cdots &0\\
	A_2A_1B_0 &A_2B_1 &\ddots &\vdots\\
	\vdots &\vdots &\ddots &0\\
	A_{N-1}\cdots A_1B_0 &A_{N-1}\cdots A_2B_1 &\cdots &B_{N-1} 
\end{bmatrix}.
\]

Furthermore, we define the block-diagonal matrices stacking the linearized constraints $C:=~\diag(C_0,\ldots,C_{N-1})$, $D:=~\diag(D_0,\ldots,D_{N-1})$,
and the stacked constraint values $\hat g_x:=~\bigl(g_x(\hat x_1),\ldots,g_x(\hat x_N)\bigr)$ and $\hat g_u:=~\bigl(g_u(\hat u_0),\ldots,g_u(\hat u_{N-1})\bigr)$.
Using the condensed dynamics \eqref{eq:linearized_condensed_dynamics}, the linearized inequality constraints can then be written compactly as
\begin{equation}
	G\Delta u\leq b+F\Delta x_0,
	\label{eq:linearized_condensed_inequalities}
\end{equation}
where $G,F,b$ collect the corresponding input, state-propagation, and offset terms.
The set of admissible initial-state perturbations induced by the linearized feasible set is
\[
\Delta\mathcal X_{\mathrm{lin}}(\hat w) := 
\left\{
\Delta x_0
\ \middle|\
\exists\,\Delta u \text{ such that }
G\Delta u\leq b+F\Delta x_0
\right\}.
\]

Let $\Delta x_0^j\in\Delta\mathcal X_{\mathrm{lin}}(\hat w)$ be the current sample and let $d^j$ be a unit search direction, i.e., $\|d^j\|_2 = 1$. 
Points on the corresponding search line are parameterized by $\Delta x_0=\Delta x_0^j+\alpha d^j$.
Substitution into \eqref{eq:linearized_condensed_inequalities} yields the boundary identification LP
\begin{maxi!}|s|[2]
	{\alpha,\Delta u}
	{\alpha}
	{\label{eq:linearized_boundary_lp}}{\Phi(\Delta x_0^j,d^j) =}
	\addConstraint{G\Delta u-\alpha Fd^j}{\leq b+F\Delta x_0^j}
	\addConstraint{\alpha}{\geq0.}
\end{maxi!}
The two boundary distances are obtained from $\alpha_+^* :=~\Phi(\Delta x_0^j,d^j)$ and $\alpha_-^* :=~\Phi(\Delta x_0^j,-d^j)$.

Starting from an initial point $\Delta x_0^1\in\Delta\mathcal X_{\mathrm{lin}}(\hat w)$, the sampling procedure is as follows. 
At iteration $j$, sample $\widetilde d^j$ uniformly from the Euclidean unit ball and set $d^j:=\widetilde d^j / \|\widetilde d^j\|_2$.
Solve \eqref{eq:linearized_boundary_lp} for $d^j$ and $-d^j$, sample $\beta^j \sim \mathcal U\bigl(-\alpha_-^*,\alpha_+^*\bigr)$,
and update $\Delta x_0^{j+1} := \Delta x_0^j+\beta^j d^j$.
The corresponding candidate initial state is $\bar x_0^{j+1} := \hat x_0+\Delta x_0^{j+1}$.
The resulting candidate \(\bar x_0^{j+1}\) belongs to \(\Xlin(\hat w)\). 
When the linearized set is an inner approximation of the nonlinear feasible set, as characterized in Proposition~\ref{prop:inner}, the candidate is already feasible for the original OCP; 
otherwise, the projection problem corrects the linearization error and returns a feasible initial state together with its corresponding optimal OCP solution.

\begin{remark}[Relation to sensitivity-based augmentation]
Sensitivity-based augmentation methods \cite{krishnamoorthy2021sensitivity,krishnamoorthy2023improved} typically prescribe a perturbation $\Delta x_0$ and compute the corresponding perturbation of the optimal trajectory from a linearized optimality system.
Here, instead, we search for $\Delta x_0$ that admit a feasible perturbation $\Delta w$ in the linearized feasible set, without requiring this perturbation to be optimal.
The nonlinear projection problem is then used to recover a feasible and locally optimal solution for the original OCP.
\end{remark}

\begin{remark}[Batched optimization]
The proposed sampling framework naturally allows for batched optimization. 
After generating multiple reference solutions through the bilevel or slack-based formulations, candidate initial states can be sampled independently from the corresponding linearized feasible sets. The resulting projection problems can then be solved simultaneously using GPU-accelerated solvers, such as MadNLP, potentially further improving the computational efficiency of data generation.
\end{remark}

\subsection{Practical sampling procedure}
\label{sec:pratical_procedure}
In practice, we combine the active-set exploration and projection-based strategies for the data-generation procedure. 
We start by generating a reference solution using either the bilevel or the slack-based formulation. 
The NMPC problem is then linearized around the reference, and $M$ candidate initial states are sampled from the corresponding linearized feasible set. 
Each candidate is used as a target in the nonlinear projection problem, which returns a feasible initial state together with a locally optimal trajectory for the corresponding original NMPC problem. 
Repeating the procedure for $N_\mathrm{p}$ reference solutions yields $N_\mathrm{p} \times M$ projected samples.
The complete sampling procedure is summarized in Algorithm~\ref{alg:combined_sampling}.
\begin{algorithm}[t]
\caption{Combined active-set exploration and projection-based sampling}
\label{alg:combined_sampling}
\begin{algorithmic}[1]
\Require Number of reference points $N_\mathrm{p}$, number of local samples $M$
\For{$i=1,\ldots,N_\mathrm{p}$}
    \State Generate a reference solution $\hat w^{(i)}$ using the bilevel \eqref{eq:bilevel} or slack-based \eqref{eq:slack_problem} formulation
    \State Linearize the NMPC problem around $\hat w^{(i)}$
    \State Construct the linearized feasible set $\Xlin(\hat w^{(i)})$
    \For{$j=1,\ldots,M$}
        \State Sample a direction in $\Delta\Xlin(\hat w^{(i)})$, identify the feasible interval by solving \eqref{eq:linearized_boundary_lp} in both directions, and draw a candidate $\bar x_0^{(i,j)}$ uniformly along this interval
		\State Solve the nonlinear projection problem \eqref{eq:projection_problem} using $\bar x_0^{(i,j)}$ as target
        \State Store the resulting feasible and locally optimal projected sample
    \EndFor
\EndFor
\State \Return $N_\mathrm{p} \times M$ projected samples
\end{algorithmic}
\end{algorithm}
Alternatively, the $M$ projection solves can be replaced by $M$ successive closed-loop MPC steps starting from each reference solution, as considered in related data-generation approaches \cite{lucia2020deep}. 
Although this also generates feasible samples under recursive feasibility, the resulting samples are concentrated along closed-loop trajectories, thus providing a less uniform exploration of the feasible set.

%% file: sections/examples.tex
\section{Numerical Examples}
\label{sec:numerical_examples}
In this section, we evaluate the proposed sampling methods on several NMPC examples, including problems with non-convex $\Xfeas$. 
We first use a low-dimensional example to illustrate Algorithm~\ref{alg:combined_sampling} under different sampling strategies, and then consider a more complex NMPC problem to assess scalability and effectiveness.
All examples are implemented in Python using CasADi~\cite{Andersson2019}. 
The bilevel active-set formulation is solved with CCOpt~\cite{pozharskiy2026ccopt}, while the projection problems and boundary-identification LPs are solved with IPOPT~\cite{wachter2006implementation}.
All experiments were run on a laptop with a 13th Gen Intel® Core™ i7-13650HX and 32 GB of memory, using one CPU core.
The code is publicly available \footnote{\url{https://github.com/piergo98/learn_mpc_data_gen.git}}.
\subsection{Discontinuous MPC}
\label{sec:discontinuous_mpc}
Consider the following nonlinear system from \cite{mpc}:
\begin{align*}
    x_{1, k+1} &= x_{1,k} + u_k, \\
    x_{2, k+1} &= x_{2,k} + u^3_k,
\end{align*}
where $x_k = (x_{1,k}, x_{2,k}) \in \mathbb{R}^2$ is the state and $u_k \in \mathbb{R}$ is the control input.
The system's optimal control law turn out to be discontinuous with a highly non-convex feasible set, making it a suitable benchmark for evaluating the proposed sampling methods.
We formulate the MPC problem \eqref{eq:ocp_explicit} with a prediction horizon of $N=2$ and stage cost $L(x_k, u_k) = \|x_k\|_2^2 + u^2_k$.
We impose the terminal constraint $x_N =~0$.
The states are constrained to the box $[-4, 4]^2$, and the input is constrained to $[-2, 2]$.
For the bilevel active-set exploration, we set $Q = I_m$, while for the projection-based sampling, we use $Q_x = 100 I_{n_x}$.
The resulting feasible set is non-convex, making this problem a suitable benchmark for evaluating the proposed sampling methods.
\begin{figure*}[t]
    \centering
    \includegraphics[width=\linewidth]{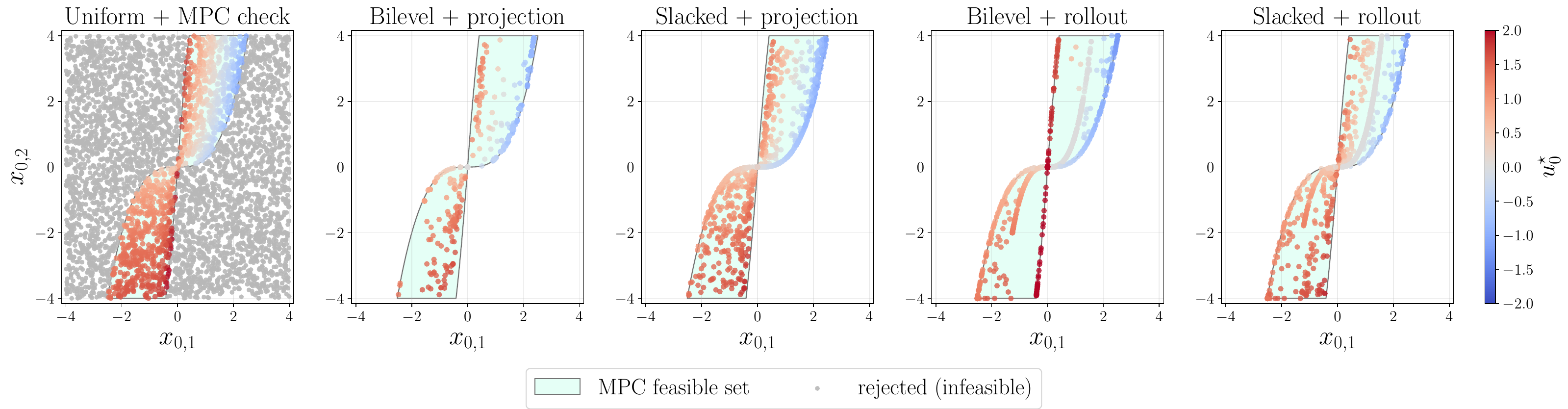}
    \caption{Initial-state samples for the five strategies shown above each panel. The shaded region is $\Xfeas$; point color indicates $u_0^*$, and gray points are infeasible.}
    \label{fig:discontinuous_mpc_comparison}
    \vspace{-0.5cm}
\end{figure*}

Here, we compare the sampled states generated using four variants: bilevel active-set exploration combined with projection-based sampling, slack-based active-set exploration combined with projection-based sampling, bilevel active-set exploration combined with closed-loop rollouts, and slack-based active-set exploration combined with closed-loop rollouts.
As a baseline, we uniformly sample from the box $[-4, 4]^2$, and solve \eqref{eq:ocp_explicit} to find feasible initial states.
For each method, optimization problems were solved until $N_\mathrm{p}\times M=1{,}000$ feasible samples were obtained, with $N_\mathrm{p}=50$ reference points and $M=20$ local samples per reference point.

Figure \ref{fig:discontinuous_mpc_comparison} displays the initial states generated by each method, along with the optimal value of the first control input $u_0$ for each initial state.

Table \ref{tab:discontinuous_mpc_comparison} summarizes the performance statistics of the five methods.
We report the number of feasible samples for the original NMPC over the entire dataset, the acceptance rate, namely, the percentage of accepted feasible samples over the total number of generated samples, and the wall-clock time required for sample generation.
The results show that the proposed active-set exploration and projection-based sampling strategies substantially improve the efficiency of feasible data generation compared with uniform sampling followed by an MPC feasibility check. 
While uniform sampling achieves an acceptance rate of only $21.3\%$, the bilevel active-set and slacked projection-based strategies reach $98.0\%$ and $100.0\%$, respectively, while all accepted samples remain feasible and satisfy the SOSC-based local-optimality check.
Although bilevel active-set exploration guarantees feasibility by construction, solver convergence failures may reduce the acceptance rate.
The active-set exploration and projection-based methods yield a $5\times$ and $14.3\times$ reduction in computational cost for the bilevel and slacked formulations, respectively, compared with uniform sampling.
The slacked closed-loop rollout strategy achieves the lowest computational cost, at the expense of a less uniform exploration of the feasible set.
Indeed, the linearized feasible-set sampling and projection-based methods provide a more uniform coverage of the feasible set, while the closed-loop rollout methods tend to cluster in certain regions.



\begin{table}[t]
\centering
\caption{Comparison of the sampling methods.}
\label{tab:discontinuous_mpc_comparison}
\begin{tabular}{lrrrr}
\toprule
\textbf{Method} &
\textbf{Acc. rate} [\%] &
\textbf{Time/Sample} [ms] \\
\midrule
Uniform
& 21.3
& 64.4 \\

Bilevel + projection
& 98.0
& 12.8 \\

Slacked + projection
& 100.0
& 4.5 \\

Bilevel + rollout
& 98.5
& 5.5 \\

Slacked + rollout
& 100.0
& 2.7 \\

\bottomrule
\end{tabular}
\vspace{-0.5cm}
\end{table}

\subsection{Nonlinear Cart-Pole}
\label{sec:cart-pole}
As a more complicated example, we consider the nonlinear cart-pole system, whose dynamics are described by the following ODEs:
\begin{align*}
    \dot{x}_1 &= x_2, \\
    \dot{x}_2 &= \frac{u + m_p l \sin(x_3) x_4^2 - m_p g \sin(x_3) \cos(x_3)}{m_c + m_p \sin^2(x_3)}, \\
    \dot{x}_3 &= x_4, \\
    \dot{x}_4 &= \frac{(m_c + m_p)g \sin(x_3) - \cos(x_3)(u + m_p l \sin(x_3)x_4^2)}{l \left( m_c + m_p \sin^2(x_3) \right)},
\end{align*}
where $x = (x_1, x_2, x_3, x_4)$ is the state vector, $u$ is the control input (force applied to the cart).
The parameters $m_c =$\qty{0.8}{\kilogram}, $m_p =$\qty{0.08}{\kilogram}, $l =$\qty{0.5}{\metre}, and $g =$\qty{9.81}{\metre/\second^2} denote the cart mass, pole mass, pole length, and gravitational acceleration, respectively.
The ODEs are discretized using a fourth-order Runge–Kutta scheme with a sampling time of \qty{0.1}{\second}. 
Accordingly, the discrete-time MPC problem \eqref{eq:ocp_explicit} is formulated with stage cost $L(x_k, u_k) = \|x_k\|_Q^2 + Ru^2_k$, and terminal cost $V_f(x_N) = \|x_N\|_E^2$, 
where $Q = \text{diag}(2.5, 0.01, 10.0, 0.01)$, $R = 0.1$, and $E$ is the solution of the continuous-time algebraic Riccati equation for the linearized system around the upright position.

Motivated by the results of Sec. \ref{sec:discontinuous_mpc}, we tested the same four sampling strategies on the cart-pole system, spanning the prediction horizon $N \in \{10,15,20,\ldots,50\}$.





\begin{figure}[b]
    \vspace{-0.5cm}
    \centering
    \includegraphics[width=0.8\linewidth]{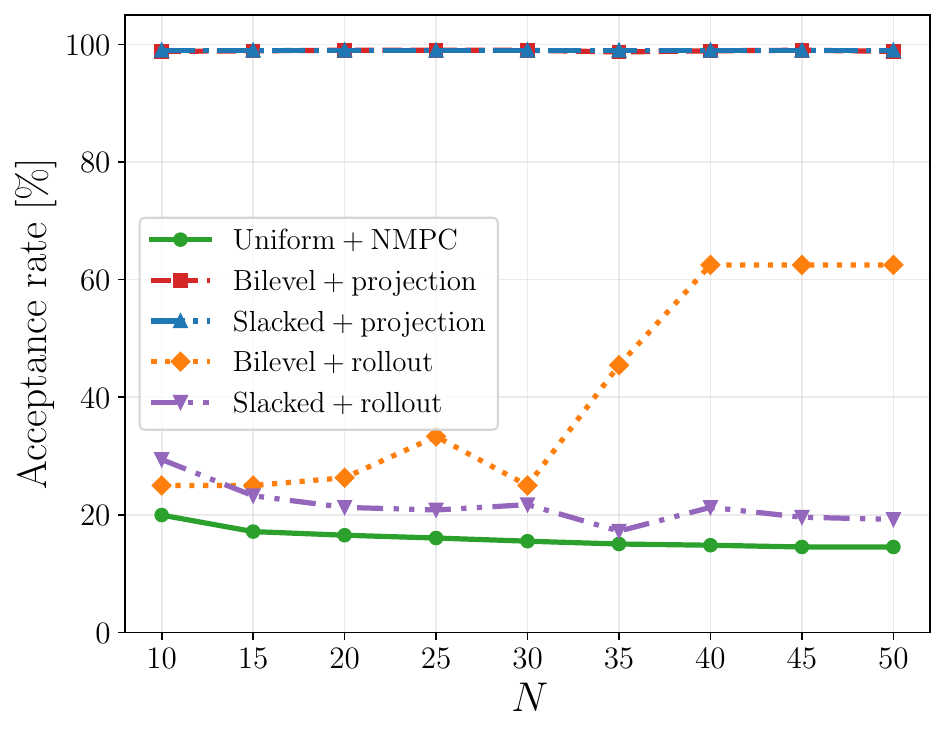}
    \caption{Acceptance rate versus prediction horizon.}
    \label{fig:acceptance_rate_horizon}
\end{figure}

\begin{figure}[t]
    \centering
    \includegraphics[width=\linewidth]{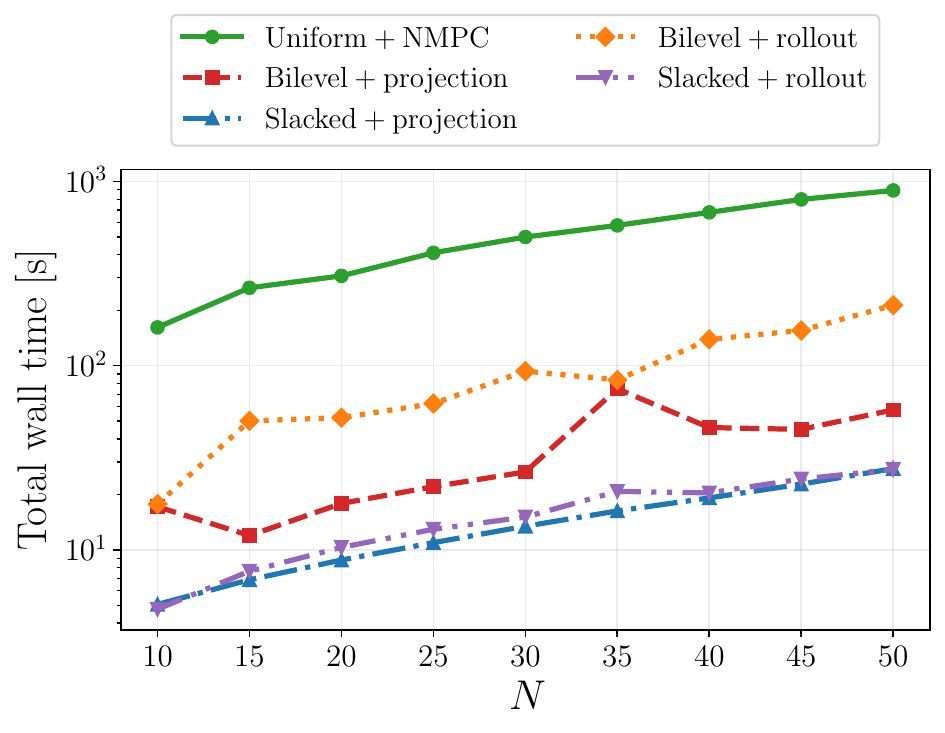}
    \caption{Total time versus prediction horizon.}
    \label{fig:total_time_horizon}
    \vspace{-0.5cm}
\end{figure}
Figure \ref{fig:acceptance_rate_horizon} and \ref{fig:total_time_horizon} show how the acceptance rate and the total time for generating $1000$ feasible samples change with the prediction horizon $N$.
Here, we chose $N_\mathrm{p}=10$ reference points and $M=100$ local samples per reference point. 

The active-set-based strategies with projection maintain a $100\%$ acceptance rate across all considered horizons, whereas uniform sampling and closed-loop rollouts exhibit lower acceptance rates, as shown in Fig. \ref{fig:acceptance_rate_horizon}.
For the rollout-based strategies, $M$ closed-loop samples are generated from each reference point; however, depending on the selected reference, some rollouts may leave the feasible region and must therefore be discarded. 
As a result, additional reference points are required to reach the target number of feasible samples, which reduces the overall acceptance rate.

Regarding the total time required to generate $1000$ feasible samples, the proposed strategies are substantially more efficient than uniform sampling, as shown in Fig.~\ref{fig:total_time_horizon}. 
The slack-based strategy combined with projection achieves the best performance, with a speed-up of approximately $37\times$ over uniform sampling for $N=25$ and $N=30$. 
Among the proposed methods, the bilevel strategy combined with closed-loop rollouts is the most computationally demanding, yet it remains approximately $4\times$ faster than uniform sampling at $N=50$, corresponding to its least favorable case. 
As expected, the computational cost of all methods increases with the prediction horizon, reflecting the growing size and complexity of the underlying NMPC problems.

%% file: sections/conclusions.tex
\section{Conclusions}
\label{sec:conclusions}
This paper proposed an efficient approach for generating informative NMPC training samples while preserving feasibility and local optimality. 
The method combines bilevel and slack-based active-set exploration with local linearized feasible-set sampling and nonlinear projection. 
Numerical results show higher acceptance rates and substantially lower computational cost than uniform sampling, with the slack-based projection strategy providing the best overall trade-off. 
The proposed approach is up to $37\times$ faster, achieves acceptance rates above $99\%$, and maintains representative feasible-set coverage, supporting its scalability for NMPC data generation.